\documentclass[aps,prl,twocolumn,showpacs,floatfix,nofootinbib]{revtex4-1}
\usepackage{amsmath,graphicx,color}
\usepackage[colorlinks,linkcolor=blue,citecolor=blue]{hyperref}

\def\P{{\mathcal P}}
\def\V{{\mathcal V}}

\def\S{{\mathcal S}}

\def\Eq#1{Eq.~(\ref{#1})}
\def\Eqs#1{Eqs.~(\ref{#1})}

\def\Fig#1{Fig.~\ref{#1}}

\def\Ref#1{Ref.~\cite{#1}}

\def\bra{\langle}
\def\ket{\rangle}

\def\sech{\mbox{sech}}
\def\jet{{\mbox{\footnotesize jet}}}
\def\hydro{{\mbox{\footnotesize hydro}}}

\def\be{\begin{equation}}
\def\ee{\end{equation}}
\def\bea{\begin{eqnarray}}
\def\eea{\end{eqnarray}}

\begin{document}

\title{Jet-medium interaction and conformal relativistic fluid dynamics}

\author{Li Yan} 
\author{Sangyong Jeon}
\author{Charles Gale}
\affiliation
{
Department of Physics,
McGill University\\
3600 rue University
Montr\'eal, QC
Canada H3A 2T8
}

\begin{abstract}
A formalism to study the mode-by-mode  response
to the energy deposition of external hard partons propagating in a relativistic fluid is developed, based on a semi-analytical  solution of  conformal  fluid-dynamics.  The soft particle production resulting from the jet-medium interaction is calculated and the recoil of the viscous medium is studied for different orientations of the relativistic jets, and for different values of the specific shear viscosity $\eta/s$.

\end{abstract}
\maketitle

\section{Introduction}

One achievement of the RHIC and LHC heavy-ion program has been the realization that the medium created at those facilities -- the quark-gluon plasma (QGP) -- could be successfully modelled theoretically using relativistic fluid dynamics. The development of hydrodynamical models that followed has further fuelled the hopes of being able to extract transport coefficients of QCD from detailed measurements of the collective behaviour of soft observables in relativistic nuclear collisions  \cite{Gale:2013da,Heinz:2013th}. 

Another remarkable property of the new state of matter created during relativistic heavy-ion collisions, the energy loss of QCD jets,  has been revealed by using ``hard probes'', i.e. probes well-calibrated with a behaviour  in vacuum that can be calculated within perturbative QCD (pQCD). An early suggestion about  using jets as a baseline in nuclear collisions  was based on elastic interactions \cite{Bjorken:1982tu} , and pictured the energy loss process as similar to the ionization loss experienced by charged particles in regular matter. It was later realized that medium-induced bremsstrahlung could be even more efficient in removing energy from the hard traveling parton (``quenching the jet'') \cite{Baier:1996kr}.   
Thanks to decades of progress in theory and in experiments, it is now firmly established that 
jet quenching in high energy heavy-ion collisions probes the 
energy loss  mechanisms 
of a hard parton traversing
a hot and dense quark-gluon plasma 
(QGP)~\cite{Connors:2017ptx,Qin:2015srf}. 
The properties of the strongly interacting medium created by colliding heavy-ion at high energies, as revealed by the analysis of jets, manifest themselves mainly through transport parameters such as the transverse momentum diffusion rate, $\hat q$, and the elastic energy loss rate $\hat e$ \cite{Baier:1996kr,Qin:2012fua}. 

How do jets affect the medium? Only recently, however, has the medium itself been included directly in jet observables. For instance, measurements at the LHC energy have
extended the reconstructed jet substructures to large jet-cone
radii~\cite{CMS:2014uca}, 
where contributions from soft
particles were found essential. 
Also, the energy imbalance observed in the 
di-hadron correlations is restored by the relatively
low transverse momentum out-of-cone
particles, which highlights  the significance of
jet-medium interaction in a theoretical description 
of jets~\cite{Chatrchyan:2011sx}.
Indeed, in a 
recent calculation of the complete jet and of its hydrodynamic
medium response~\cite{Tachibana:2017syd}, the theoretical interpretation of the measured jet shape requires the jet-induced
medium excitations to be included in the dynamical evolution of the fluid background. Even though a growing number of efforts have been devoted to 
simulate jet partons propagating through
a QGP medium 
(cf.~\cite{CasalderreySolana:2004qm,Chesler:2007sv,Stoecker:2007su,Neufeld:2008fi,Tachibana:2014lja,Qin:2009uh,Betz:2008ka,Chaudhuri:2005vc,Shuryak:2013cja,Chen:2017zte,Milhano:2017nzm,Floerchinger:2014yqa}), a wholistic treatment which includes also the medium recoil remains challenging. Some of the complications seeking resolution  stem from the dynamically-evolving  medium. For example, the medium expansion will  
distort the generated conical flow structure expected in a 
static system~\cite{CasalderreySolana:2004qm}. 

In light of the complications involved in a complete and consistent theoretical  treatment in 3D involving jets and fluid background, we turn to an approach with a simpler geometry, but which has a formal solution. In this work we use the flow solution put forward by Gubser and collaborator~\cite{Gubser:2010ze,Gubser:2010ui},
to develop a semi-analytical formalism for the treatment of the jet-medium 
interaction on a mode-by-mode basis.
In this way, perturbation modes of small wave
numbers are captured in the linearized 
hydrodynamic equation of motion,
while modes associated with large wave numbers
can be safely ignored since they are stronlgy suppressed
by viscosity.

To be more specific, 
we consider the 
energy-momentum conservation of the jet parton and
fluid system,
\be
\partial_\mu T^{\mu\nu} = \partial_\mu(
T^{\mu\nu}_\hydro+T^{\mu\nu}_\jet+\delta T^{\mu\nu})=0\,,
\ee
where $T^{\mu\nu}_\hydro$ and $T^{\mu\nu}_\jet$ are 
energy-momentum tensor for the background fluid
and the jet parton, respectively. Effect of jet-medium
interaction is described by $\delta T^{\mu\nu}$. 
For the  perturbations induced by the jet parton,
the energy-momentum conservation of the whole system
can be separated into the 
equation of motion of the
background fluid $\partial_\mu T_\hydro^{\mu\nu}=0$, 
 and a linearized 
equation of motion for the 
jet-medium interaction~\cite{CasalderreySolana:2004qm},
\begin{align}
\label{eq:eom}
\partial_\mu \delta T^{\mu\nu}=-\partial_\mu T^{\mu\nu}_\jet=
J^\nu\,,
\end{align}
where we have written effectively the loss of energy-momentum from
the parton as a source current.
To solve \Eq{eq:eom} with respect to 
the source current for jet-medium interaction,
one needs explicit form of $\delta T^{\mu\nu}$, which however
is not given \emph{a priori}. It is only for the long-wavelength
modes (with wave number $k$ satisfying $k\lambda_{\mbox{\tiny mfp}}\ll 1$),
can one identify $\delta \tilde T^{\mu\nu}(k)=\delta \tilde T^{\mu\nu}_\hydro(k)$~\cite{Iancu:2015uja}
in terms of perturbations in the energy density, pressure,
and flow velocity.
In a mode-by-mode analysis as in this work, 
this is realized naturally by focusing on long-wavelength modes~\cite{Shuryak:2013cja,CasalderreySolana:2004qm,Chesler:2007sv}. 

Inspired by \Ref{Tachibana:2017syd}, by applying the Landau matching
condition to the evolution of jet parton distribution function we find
the following source current in \Eq{eq:eom} for a light-like
parton ($|{\bf v}_\jet|=c$)~\cite{Singh:2014uca},
\be
\label{eq:source}
J^\mu(t,{\bf x})=\hat e v^\mu_\jet n_\jet(t,{\bf x})\,,
\ee
where $\hat e=\bra \Delta E\ket/dt$ is the average rate of energy loss 
of the jet parton, and $v^\mu_\jet=(1,{\bf v}_\jet)$ is the parton four-velocity. The density of jet partons $n_\jet$ contains information on
the source
shape.  
We consider a boost-invariant configuration in this work. Especially, 
the jet parton
passing through the transverse plane 
presents a ``knife-shape'', which spans the whole
rapidity range and is captured by the density as
$n_\jet(t,{\bf x}_\perp)=\delta^{(2)}({\bf x}_\perp - {\bf v}_{\jet\perp}\Delta \tau)/\tau$,
where $\tau$ is the longitudinal proper time. 
The boost-invariant assumption of a jet parton is an idealization, but it captures qualitatively the main effect of an energetic parton going
through the QGP medium. In the context of this work, a boost-invariant structure corresponds quantitatively to the
mode with the longest wavelength along the space-time rapidity, whose dynamical evolution
is more sensitive in the jet-medium interaction.
In obtaining the source \Eq{eq:source}, we have also assumed that the 
jet parton is of sufficiently high energy, $E_\jet\gg T$, so that
contribution from parton transverse momentum broadening 
is negligible.

\section{Mode-by-mode hydrodynamics}

Solutions to fluid-dynamical equations are few and far in between, therefore one may gain considerable insight from cases that are exactly solvable. 
The method developed by Gubser and Yarom for solving the equations of viscous hydrodynamics \emph{analytically}
characterizes the 
longitudinal and radial
expansions of a conformal
fluid system ($e=3\P$), and incorporates rotational symmetry in azimuth and
Bjorken boosts in the longitudinal direction
~\cite{Gubser:2010ze,Gubser:2010ui}. Importantly, those solutions are regularly used to test modern hydrodynamics codes for accuracy \cite{Marrochio:2013wla,Denicol:2014xca,deSouza:2015ena,Okamoto:2017ukz}.  

The solution technique consists of making a coordinate transformation from
Milne space-time $(\tau, r, \phi, \xi)$, where $\tau$ is 
proper time and $\xi$ is the space-time rapidity,
to a $dS_3\times R$ coordinate system $(\rho, \theta, \phi, \xi)$, 
through
\begin{subequations}
\label{eq:transf_cor}
\begin{align}
\sinh\rho=&-\frac{1-q^2\tau^2+q^2r^2}{2q\tau}\,,\\
\tan\theta=&\frac{2qr}{1+q^2\tau^2-q^2r^2}\,,
\end{align}
\end{subequations}
so that a $SO(3)$ rotational symmetry becomes manifest
in the subspace $(\theta,\phi)$, for the transformed metric tensor. The parameter 
$q$ in \Eq{eq:transf_cor}
specifies the inverse length scale of the system.

The background medium evolution
has an analytical solution in this new coordinate system. 
In what follows, we shall indicate hydro variables in this
new  system with an overbar.
While the flow four velocity is explicitly determined as
$\bar u^\mu=(1,0,0,0)$, with first order viscous corrections (Navier-Stokes hydro) the 
energy density is solved as~\cite{Gubser:2010ze}
$
\label{eq:en}
\bar \epsilon(\rho)=
(\cosh\rho)^{-8/3}\left[
\bar T_0+H_0 F_d(\rho)/3
\right]^4\,,
$
where $F_d(\rho)$ is an analytical function whose form is given in \Ref{Gubser:2010ze}.
In the expression of energy density,
$\bar T_0$ and $H_0$ are constant parameters to be fixed by the system multiplicity
and shear viscosity over entropy ratio, respectively.
Owing to symmetry constraints, the 
energy density depends only on 
the de Sitter time $\rho$. 
Solutions with respect to second order viscous corrections
of a conformal fluid~\cite{Baier:2007ix}
can be achieved by solving an ordinary differential equation.
Hydro variables in the original
Milne space-time and in the $dS_3\times R$ frame are related
to each other through mappings associated with
\Eq{eq:transf_cor}. In particular, the source term in \Eq{eq:source}
in the $dS_3\times R$ frame
becomes 
$
\bar J_\mu = \hat e \,\bar n_\jet\, \bar v_\mu^\jet 
$ with, 
$\bar n_\jet=\delta(\theta-\theta(\rho))\delta(\phi-\phi(\rho))/\cosh^2\rho\sin\theta$.

Since transverse coordinates of the Milne space-time
possess apparent rotational symmetry in terms of $(\theta,\phi)$, the 
mode decomposition of \Eq{eq:eom} can be achieved using spherical
harmonics, resulting in 
perturbations of temperature 
and
flow velocity: 
\begin{subequations}
\label{eq:modes}
\begin{align}
\delta \bar T =& \bar T\sum_{lm}\int\frac{dk_\xi}{2\pi} t^{lm}(\rho) Y_{lm}(\theta,\phi) e^{ik_\xi \xi}\\
\delta \bar u_i =& \sum_{lm}\int\frac{dk_\xi}{2\pi}\left[v_s^{lm}(\rho)\Psi^{lm}_i(\theta,\phi) 
+ v_v^{lm}(\rho)\Phi_i^{lm}(\theta, \phi)\right]e^{ik_\xi \xi}\\
\delta \bar u_\xi =& \sum_{lm}\int \frac{dk_\xi}{2\pi}v_\xi^{lm}(\rho) Y_{lm}(\theta,\phi)e^{ik_\xi \xi}\,.
\end{align}
\end{subequations}
Here, $Y_{lm}$ is the scalar spherical harmonics, while
$\Psi_i^{lm}$ and $\Phi_i^{lm}$ are vector spherical 
harmonics which have a vanishing curl and divergence 
in the subspace $(\theta,\phi)$, respectively~\cite{0143-0807-6-4-014}. The quantity 
$k_\xi$ is the conjugate coordinate of the space-time rapidity $\xi$.
Mode decomposition in \Eqs{eq:modes} results in three scalar modes 
$(t,v_s,v_\xi)$ and
one vector mode $v_v$.
Similarly, one also has the mode decomposition of the source current
$\bar J_\mu$, which leads to a set of  
\begin{subequations}
\begin{align}
\bar J_\rho = &\sum_{lm}\int\frac{dk_\xi}{2\pi} c^{lm}_\rho Y_{lm}(\theta,\phi) e^{i k_\xi \xi}\\
\bar J_i = & \sum_{lm}\int\frac{dk_\xi}{2\pi}[c^{lm}_s \Psi_i^{lm}(\theta,\phi) + c^{lm}_v \Phi_i^{lm}(\theta,\phi)]e^{i k_\xi \xi}\\
\bar J_\xi =&\sum_{lm}\int\frac{dk_\xi}{2\pi} c^{lm}_\xi Y_{lm}(\theta,\phi)e^{i k_\xi \xi}\,.
\end{align}
\end{subequations}

The typical length scale of each mode in the decomposition
is determined by index $(l,m)$ and
$k_\xi$. In particular in this boost-invariant case, in which
only the modes associated with $k_\xi=0$ contribute,
we found it 
mostly dependent on $l$, 
and behaves like
$\sim1/( \sqrt{l}q/3)$.\footnote{
This is inferred empirically from the width at half maximum 
of the basis function of
the mode decomposition.
}
>From this, one can deduce that for a mode to be considered
hydrodynamic, it must satisfy
$\lambda_{\mbox{\tiny mfp}}\sqrt{l}q/3 \sim \eta\sqrt{l}q/3sT\ll1$.
For the fluid medium created in high energy heavy-ion collisions, 
this inequality is usually satisfied up to a typical value of $l\ll 10^2$. 

In terms of these scalar and vector modes 
$\bar \V^{lm}=(t^{lm},v^{lm}_s,v^{lm}_\xi,v^{lm}_v)$,
the  linearized hydrodynamics \Eq{eq:eom} 
reduces to a set of coupled differential equations 
\be
\label{eq:jet_eom}
\partial_\rho\bar \V^{lm}(\rho,k_\xi) = 
-\Gamma(\rho,l,k_\xi)\bar \V^{lm}(\rho,k_\xi) + \bar \S^{lm}(\rho,k_\xi)\,,
\ee
where $\Gamma$, 
whose explicitly expression can be found in Eq. (109) of~\Ref{Gubser:2010ze},
is a matrix determined by the background medium
expansion. The source term $\bar\S^{lm}$ is 
\be
\bar\S^{lm}=
 \begin{pmatrix}
  -\frac{1}{3\bar w}c_\rho^{lm}\\
   -\frac{2\bar T \tanh \rho}{3\bar w\bar T'} c_s^{lm}\\
  \frac{\bar T}{\bar w(\bar T+H_0 \tanh\rho)}c_\xi^{lm} \\
  -\frac{2\bar T \tanh \rho}{3\bar w\bar T'}c_v^{lm}
 \end{pmatrix}\,,
\ee
where $\bar w=\bar e+\bar \P$ is the enthalpy density.

The matrix $\Gamma(\rho,l,k_\xi)$ is 
block-diagonalized in such a way that the scalar modes and the vector modes
decouple.
Furthermore, $v_\xi$ is also decoupled from the 
rest of the scalar modes when $k_\xi=0$, corresponding to a 
system including also boost-invariant hydro perturbations.
Thus we shall ignore the contribution of $v_\xi$ mode in what
follows. 
In the case of an ideal fluid with $H_0=0$, 
$\Gamma$
has two scalar eigenmodes with eigenvalues 
\be
\label{eq:eigen}
\lambda_{\pm}=-\frac{1}{3}\tanh\rho\pm \frac{1}{3}\sech\rho
\sqrt{\sinh^2\rho-3l(l+1)}\,.
\ee
When $\sinh^2\rho<3l(l+1)$, 
$\lambda_{\pm}$ becomes complex, indicating 
sound wave propagation of the scalar modes. 
In the original Milne space-time, complex eigenvalues of scalar
modes appear mostly at late $\tau$, while at very early
$\tau$, scalar modes are still diffusive.
For the vector mode, however,
$\Gamma_v=-\frac{2}{3}\tanh\rho$ is always real, 
which is purely diffusive. The viscous corrections 
damp mode evolution and the damping is systematically
stronger for higher order modes (larger $l$)~\cite{Gubser:2010ui}.

\section{Jet-medium interaction in Heavy-ion collisions}

\begin{figure}
\begin{center}
\includegraphics[width=0.25\textwidth] {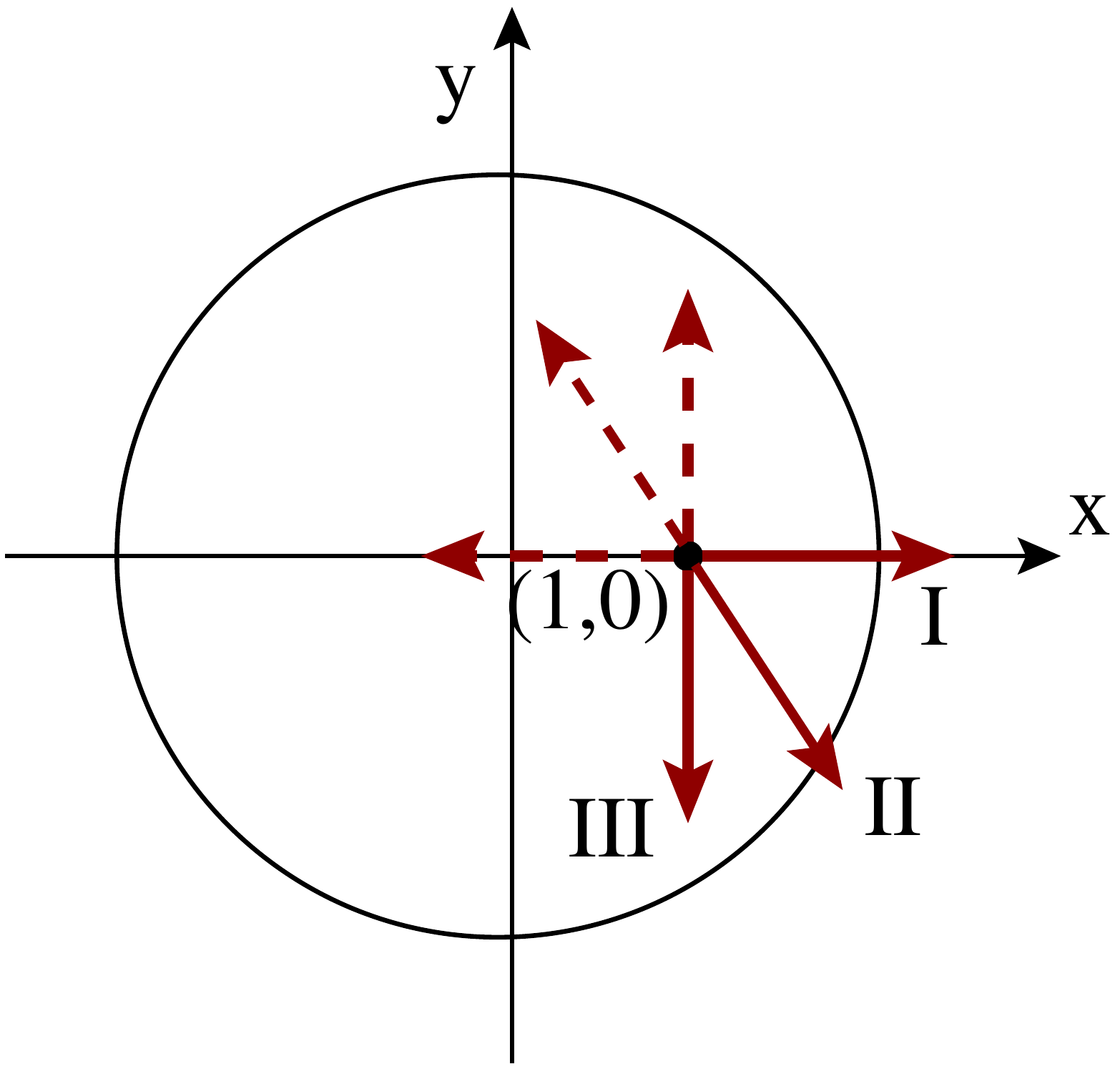}
\caption{
\label{fig:cases}(Color online)
Three events with di-jet considered in this work. Solid 
arrows indicate jet partons generating the near-side (leading) peak
of the observed spectrum, while dashed arrows are
those generating the away-side (sub-leading) peak. 
}
\end{center}
\end{figure}

We solve \Eq{eq:jet_eom} for 
ultra-central Pb+Pb collisions at the LHC energy $\sqrt{s_{NN}}=2.76$ TeV.
This is done in the semi-analytical solution of conformal viscous hydrodynamics 
by specifying $\bar T_0=7.3$ and $1/q=4.3$ fm~\cite{Staig:2011wj,Yan:2015lfa}.
To stress both the effects of medium expansion and
dissipation, 
we consider three representative 
events with one pair of boost-invariant back-to-back jet partons
($k_\xi=0$). These events are illustrated in \Fig{fig:cases},
corresponding respectively to back-to-back partons 
oriented
along $x$-axis (case I), at an angle of $\pi/3$ (case II),  
and at an angle of $\pi/2$ (case III), all 
starting from $\vec x^{\jet}_{\perp}=
(1.0,0)$ at $\tau_0=0.5$ fm/c. 
For each of these cases, 
we calculate separately the medium response
to the near-side and away-side jet partons, whose superposition results in 
the medium response to a di-jet in linearized hydrodynamics.  It is 
worth mentioning that 
each individual jet parton may as well be recognized as a parton in 
a $\gamma$-jet, for which the QGP medium is transparent to the
photon recoiling against a parton.  

We adopt in this work a $T^2$-dependent jet energy loss 
rate: $\hat e=\kappa T^2$~\cite{Casalderrey-Solana:2014bpa,Ficnar:2013qxa,Betz:2011tu},
while generalization to other parameterizations is straightforward.
We can determine the coefficient $\kappa$ via the dissipative properties
of the QGP medium. 
In a weakly coupled system,
where medium dynamical properties can be estimated perturbatively 
based on a quasiparticle assumption, 
the coefficient is inversely proportional to the specific 
viscosity 
as $\kappa\approx s/3\eta$, 
from the proposed relation $1.25\, T^3/\hat q\approx \,\eta/s$~\cite{Majumder:2007zh}
and the fluctuation-dissipation relation $\hat q=4\hat eT$~\cite{Moore:2004tg,Qin:2009gw}.
As a consequence, for a weakly coupled system, 
a jet parton loses less energy to a more viscous medium
and 
one expects
correspondingly a suppression of the
jet-medium interaction inversely proportional to $\eta/s$, 
an effect we refer to 
as \emph{dynamical} viscous suppression.
Whereas for a strongly coupled system, there is no obvious relation between
the coefficient $\kappa$ and $\eta/s$, except 
a lower bound to the rate of jet energy
loss, $\kappa\gg s/3\eta$, 
according to $1.25\,T^3/\hat q\ll \eta/s$~\cite{Majumder:2007zh},
the effect of \emph{dynamical} viscous suppression
is less clear. 
Nevertheless, considering some extreme case of strongly coupled QCD medium 
where jet energy loss has very little dependence on the medium
dissipative properties, \emph{dynamical} viscous suppression
may be negligible. We shall return to this point later.

One may now proceed to study  
the medium response to a jet parton, and to also explore 
the effect of shear viscosity on that system. 
Let us first 
start with the weakly coupled system paradigm (which relates $\hat q$ with $\eta/s$), with
the values of the specific shear viscosity being considered as 
$\eta/s=1/4\pi$, and $2/4\pi$ \footnote{In this work we shall take $\eta/s=1/4\pi$  as 
the specific shear viscosity in a ``weakly coupled'' system, 
even though it is known to be the default specific shear viscosity for systems 
analyzed with AdS/CFT techniques \cite{Kovtun:2004de}.}. 
We have verified that, with these values, 
the mode summation always
converged up to mode $l<35$,
so that a reliable solution of the jet-medium interaction
was achieved for the viscous fluid.
{For instance, the additional contribution to the energy deposited from the jet parton
becomes negeligible for $l \ge 35$.
A detailed comparison of the present result with complete numerical hydro simulations of the
jet-medium interaction in Milne space-time  will be given in \Ref{Singh:2014uca}. 

\begin{figure}
\begin{center}
\includegraphics[width=0.4\textwidth] {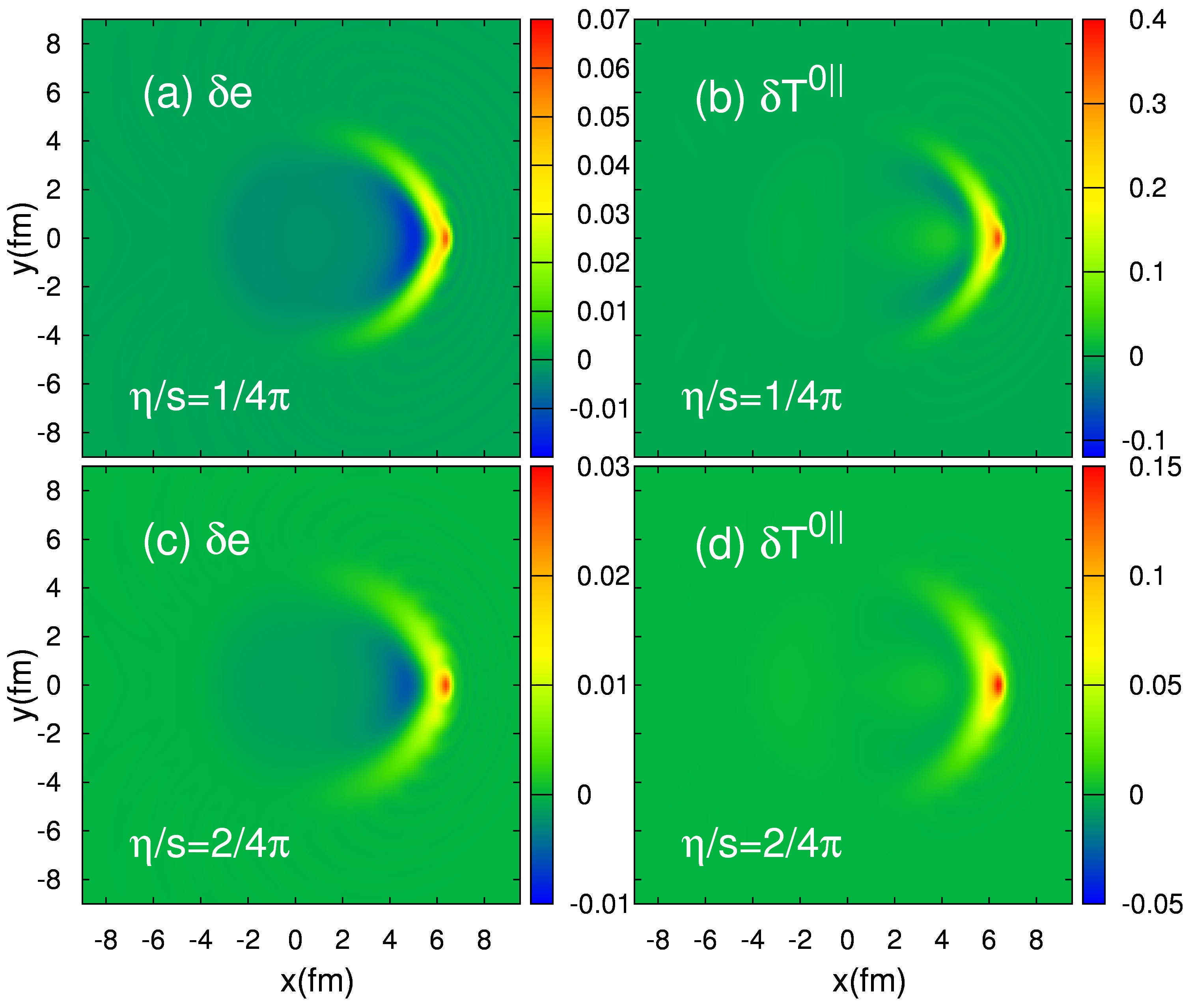}
\caption{
\label{fig:den}(Color online)
Medium excitations on top of expanding viscous systems with 
$\eta/s=1/4\pi$ (upper rows)
and $\eta/s=2/4\pi$ (lower rows) with 
respect to the near-side jet parton of event I, 
plotted in terms of energy density $\delta e$ (left panels)
and energy flux 
$\delta T^{0\parallel}$ (right panels),
at $\tau=6.0$ fm/c. The colour coding reflects units of GeV/fm$^3$. 
}
\end{center}
\end{figure}

As a supersonic object, a 
light-like jet parton going through QGP creates
a conical flow
~\cite{CasalderreySolana:2004qm}. 
For a static medium, the
conical flow has been found as 
a consequence of coherent superposition of sound wave
propagation, with the cone angle 
$\theta^M_0$
related to the speed of sound $c_s$:
$\theta^M_0=2\sin^{-1}(c_s/c)\approx 70^o$~\cite{Chesler:2007sv,CasalderreySolana:2004qm}.
When the system expands, sound propagation gets Lorentz boosted 
by the radial flow, which effectively distorts the cone structure. 
For instance, 
when a jet parton goes with the medium expansion, as the
trigger parton considered in our event I, the medium expansion
``pushes" the sound waves outwards, leading to a cone 
angle larger than $\theta^M_0$. This is seen 
in
\Fig{fig:den} (a) and (c), 
where 
the induced medium response 
on top of an expanding medium, 
is presented in terms of perturbations of 
energy density $\delta e$, 
at $\tau=6$ fm/c\;\footnote{
At $\tau=6$ fm/c, the medium cools down to a temperature
$T\lesssim 130$ MeV, at which fluid systems in
heavy-ion collisions are commonly considered to
freeze-out. See later discussions for more details.
}. 
Despite a larger cone angle, 
the formed Mach cone exhibits  features very 
similar to those observed in a static medium~\cite{Chesler:2007sv,Neufeld:2008fi}. 
In particular,
the depletion behind the cone structure  responsible for the 
sonic boom is observed. 
Sound modes also contribute
to energy flux of the excited medium,
which explains a similar Mach cone structure
in $\delta T^{0\parallel}$ in \Fig{fig:den} (b) and (d).
Note that in $\delta T^{0\parallel}$,
 a diffusive 
wake is generated behind the
shock, which contains excited kinetic energy 
flowing along the jet parton. 
However, this diffusive wake 
is \emph{not} visible in $\delta e$~\cite{Chesler:2007sv}.

For a weakly-coupled system, one observes that the viscous 
effects on the jet-medium interaction
are many. First, 
an overall reduction of the hydro excitations in the medium
response is expected due to the \emph{dynamical} viscous
suppression,
which
is inversely proportional to $\eta/s$. 
In addition, evolution of hydro modes
is further damped by shear viscosity.
Higher order modes get stronger relative viscous corrections~\cite{Schenke:2011bn},
which is roughly proportional to $\exp(-\Delta tk^2\eta/sT)$ with $k$ the wave number
(In the Gubser solution, this factor corresponds to $\exp(-l^2 H_0\Delta \rho)$). Thereby, for each mode there is a suppression factor $\exp(-\Delta tk^2\eta/sT)/(\eta/s)$.
As a consequence, comparing 
\Fig{fig:den} (c) and (d) to
\Fig{fig:den} (a) and (b), although a Mach cone is 
formed 
with its angle barely affected,\footnote{
This can be understood since shear viscosity in the conformal
fluid system does not significantly 
change the speed of sound $c_s$
nor the medium expansion.  
}
the shock wave is smeared with its amplitude reduced. 
Apart from the reduction due to \emph{dynamical} viscous suppression 
(by exactly a factor of 2 when increasing $\eta/s$
from $1/4\pi$ to $2/4\pi$), additional reduction and 
smearing of the cone structure 
is entirely a fluidity effect reflected as 
the shock waves decaying and spreading
 some distance from the vertex, which
we may refer to as the \emph{hydro} viscous suppression.
One should note that increasing specific viscosity from $1/4\pi$ to
$2/4\pi$, the induced reduction of conical flow is 
dominated by the \emph{dynamical} viscous suppression.
In a similar way, in \Fig{fig:den} (b) and (d), the diffusive wake
in $\delta T^{0\parallel}$ gets broadened 
due to medium viscous corrections.

\begin{figure}
\begin{center}
\includegraphics[width=0.45\textwidth] {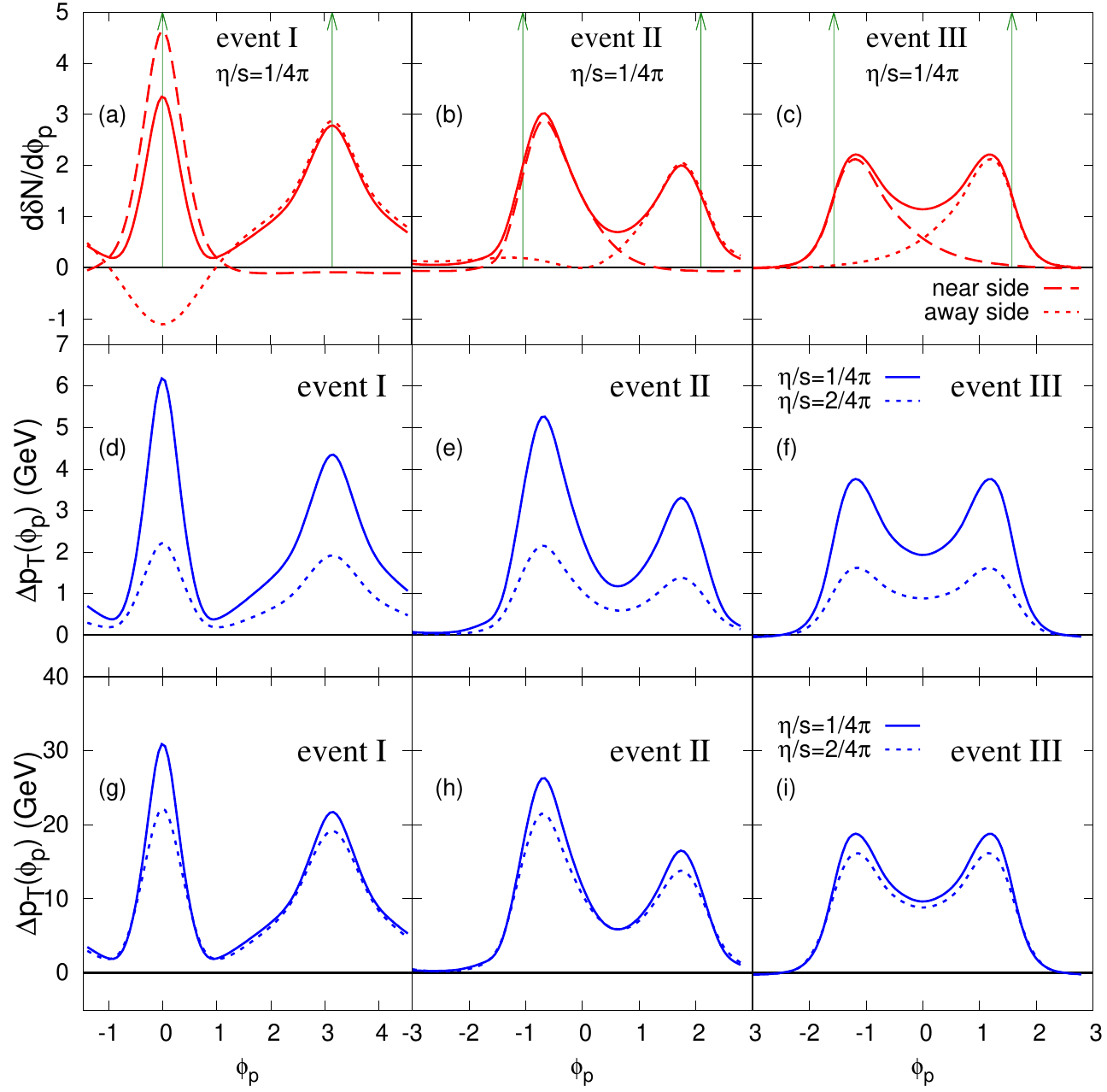}
\caption{
\label{fig:spec} (Color online)
First row: pion number density generated
from the medium response to jet partons as a function of
azimuthal angle. The near- and away-side number density are shown separately, as well as their sum (solid line). Second row: transverse energy of the induced
pions from the medium response to a di-jet 
in a weakly-coupled system. 
Third row: transverse energy of the induced
pions from jet-medium interaction for a strongly-coupled 
system. 
See the main text for more details. 
A lower cut of $p_T\ge1$ GeV has been applied 
in these plots.
}
\end{center}
\end{figure}

The excited medium response to the jet partons particlizes
when it is decoupled from the fluid dynamical
evolution. We follow 
the standard Cooper-Frye freeze-out prescription \cite{Cooper:1974mv} at a constant
proper time $\tau_f=6$ fm/c, 
\be
\label{eq:freeze}
E\frac{d\delta N}{d^3 p}=
\int d \Sigma_\mu p^\mu \delta f\,,
\ee
to compute the 
contributions to the particle spectrum from jet-medium
interaction. 
In \Eq{eq:freeze}, $\delta f$ is obtained
from the difference
between a system with and without external jet source, 
and includes the appropriate viscous correction.
For simplicity, we only consider pion production from the 
freeze-out surface and ignore further interactions of hadrons.

In \Fig{fig:spec} (a), (b) and (c), the generated
pion number density of the near-side (dashed lines) 
and away-side (dotted lines) from jet-medium interaction 
are plotted separately 
as a function of azimuthal angle $\phi_p$,
with medium specific viscosity $\eta/s=1/4\pi$. 
The green arrows indicate the directions of external jet
partons.
Medium response to an individual parton leads to a peak
of width of order $O$(1)
in the associated particle spectrum. 
Note that the width in azimuth 
identifies with jet cone size in the
boost-invariant configuration.
The overall height of the peak
is related to the total energy loss of the parton, which
is further determined by the jet parton's path. The shape 
of the peak reflects the structure of conical flow.
Superposition of the near-side and away-side gives rise to a 
double-peak distribution (solid lines) 
of the produced pions from a di-jet. As a consequence of
medium expansion, in event II and III, 
we notice that the two centers of 
the double-peak are shifted from the original directions 
back-to-back partons. 
It is also worth mentioning that the depletion in the 
Mach cone results in a depletion in the particle spectrum
in the opposite direction of the jet parton, as is evident in 
\Fig{fig:spec} (a). 

Viscous effect on the associated particle spectrum is 
revealed in \Fig{fig:spec} (d), (e) and (f), where we present
the total transverse energy induced from the medium 
response with respect to a di-jet, for $\eta/s=1/4\pi$
and $2/4\pi$ respectively. 
Shear viscosity affects the calculated particle spectrum in two 
ways. First, it smears and suppresses 
the induced Mach cone structure, as a combined 
consequence of the \emph{dynamical} and \emph{hydro}
viscous suppressions. 
Second, it modifies the phase space distribution 
in the Cooper-Frye freeze-out. By switching on and
off the viscous corrections at freeze-out, we find
the latter one is actually minor. Therefore, 
one observes in \Fig{fig:spec} (d), (e) and (f) the strong
suppression of the double-peak structure due to viscosity,
which again is mostly due to the \emph{dynamical} viscous suppression. 
We have verified that, without introducing the dynamical viscous suppression, the change in the particle spectrum in the weakly-coupled system 
is small.

At last, let us briefly discuss the jet-medium 
interaction in a strongly coupled
system with a negligible \emph{dynamical}
viscous suppression, with a $\eta/s$-independent 
jet energy loss.
As a crude estimate,
we consider a 
value of $\kappa=20\pi/3$, so that $\kappa\gg 3s/\eta$ is
roughly satisfied with respect 
to $\eta/s=1/4\pi$ and $2/4\pi$.
Results are shown in
\Fig{fig:spec} (g), (h) and (i). 
Despite the fact our $T^2$-dependent jet energy loss rate does not 
rigorously implement a parameterization for the strongly coupled 
systems, the 
absence of \emph{dynamical} viscous suppression is the
key factor in the analysis. 
Since the only
viscous effect is \emph{hydro} viscous suppression,
the changes in the spectrum are entirely from 
viscous damping of hydro modes. 
As anticipated,
short-scale modes get stronger viscous damping, which explains the 
stronger suppression around the peaks. However, the
overall viscous corrections to the double-peak structure are not large. Our analysis shows however that the details of the jet energy-loss mechanism do leave a potentially measurable imprint on the spectra of soft produced particles. 

\section{Summary and conclusions}

We have developed a formalism of solving hydrodynamical 
response to an external source mode-by-mode, based on
the Gubser's solution to conformal fluid systems. 
With respect to
the ultra-central Pb+Pb collisions at the LHC energy,
the medium response to a light-like jet parton is analyzed,
in which a conical flow structure is observed. 
The structure
of the cone receives modifications
from both the medium expansion and viscous damping, which 
then propagates to the generated particle spectrum.
After Cooper-Frye freeze-out, we observe 
that a relatively wide 
peak structure in the associated particle spectrum
is generated from the conical flow.

Viscous effect on the jet-medium interaction can be 
revealed in the suppressed peak structure of the particle
spectrum. It is a result of the \emph{dynamical}
viscous suppression which accounts for a 
reduced jet energy loss rate in a more dissipative fluid,
and \emph{hydro} viscous suppression which accounts
for medium viscous damping of hydro modes.  
With respect to the peak structure of the particle spectrum,
we found that the dominant suppression is from
the \emph{dynamical} viscous suppression, 
while \emph{hydro} viscous suppression is less important,
which was clearly demonstrated in the extreme case of strongly
coupled system.
In a weakly coupled system, where \emph{dynamical}
viscous suppression is expected from the well-established 
relation between
jet parton energy loss and $\eta/s$,
viscous effect on the jet-medium interaction is strong.
Since \emph{dynamical} viscous suppression is inversely
proportional to $\eta/s$, it 
implies a novel measure of medium transport
coefficients in the observed jet substructures. 


This present work concerns boost-invariant configuration of the background 
medium and the jet parton shape. Correspondingly, the 
resulting particle spectrum (as what is shown in Fig. (\ref{fig:spec})) is obtained 
under the condition $\Delta \eta=0$. In terms of hydro modes, that case is associated with
the mode with longest wave-length in the space-time rapidity, or $k_\xi=0$. For
more realistic situations, all higher order $k_\xi$ modes should be taken into account
in the mode summation.  But for each of the $k_\xi$ mode, we expect similarly that
the effect from 
the \emph{dynamical} viscous suppression will dominate over that from the \emph{hydrodynamical}
viscous suppression. Therefore, even after the summation over $k_\xi$ modes, or to say,
for a realistic jet parton localized in space-time rapidity,
the conclusion that the dominant viscous effect  to the induced jet-medium interaction 
is the \emph{dynamical}
viscous suppression is robust.

\section*{Acknowledgements}
We thank D. Pablos for helpful discussions.
This work was supported in part by the Natural Sciences and Engineering Research Council of Canada. C. G. gratefully acknowledges support from the Canada Council for the Arts through its Killam Research Fellowship program.

\bibliography{jet_gub.bib}

\end{document}